\documentclass[aps,prl,showpacs,reprint]{revtex4-1} 
\usepackage{graphics}
\usepackage{graphicx,epsfig}
\usepackage{amsmath}
\usepackage{amssymb}
\usepackage{amsthm}
\usepackage{exscale}
\usepackage[mathscr]{eucal}
\usepackage{bm}
\usepackage[dvipsnames]{color}
\usepackage{url}
\usepackage[perpage]{footmisc}
\usepackage{tabularx}
\usepackage{array}
\usepackage{natbib}
\usepackage{float}
\usepackage[section]{placeins}
\usepackage[version=3]{mhchem} 
\graphicspath{{Figures/EPS/}{Figures/}}

\begin{document}
\author{Wei Chen$^{1,2}$, Adam S. Foster$^{1}$, Mikko J. Alava$^{1}$, and Lasse Laurson$^{1}$}
\email{lasse.laurson@aalto.fi}
\affiliation{$^1$COMP Centre of Excellence, Department of Applied Physics,\\
Aalto University, P.O. Box 11100, 00076 Aalto, Espoo, Finland}
\affiliation{$^2$Supercomputing Center of CAS,  Computer Network Information Center, 
Chinese Academy of Sciences, Beijing 100190, China}

\title{Stick-Slip Control in Nanoscale Boundary Lubrication by Surface Wettability}

\begin{abstract}
We study the effect of atomic scale surface-lubricant interactions on nanoscale 
boundary-lubricated friction, by considering two example surfaces - hydrophilic mica 
and hydrophobic graphene - confining thin layers of water in molecular dynamics 
simulations. We observe stick-slip dynamics for thin water films confined by mica 
sheets, involving periodic breaking-reforming transitions of atomic scale capillary 
water bridges formed around the potassium ions of mica. However, only smooth sliding 
without stick-slip events is observed for water confined by graphene, as well as for 
thicker water layers confined by mica. Thus, our results illustrate how atomic scale
details affect the wettability of the confining surfaces, and consequently control
the presence or absence of stick-slip dynamics in nanoscale friction.
\end{abstract}

\maketitle

Understanding friction plays a central role in technological applications and phenomena in diverse
fields ranging from micromechanical devices to bioengineering \cite{PhysRevLett.97.076103} and to 
earthquakes \cite{scholz1998earthquakes}. Given the continuing miniaturization of 
mechanical devices towards the nanoscale \cite{bhushan1995nanotribology}, improved understanding of 
friction and wear could help in reducing energy consumption, improving reliability and 
extending service life. Indeed, an important part of their design process 
consists of trying to minimize friction and to eliminate stick-slip dynamics 
\cite{dieterich1978time}. 

Stick-slip control in lubricated friction is of particular importance given the vast amount
of applications where lubricants are used to reduce the detrimental effects of friction 
and wear \cite{hamrock2004fundamentals}. Examples of mechanisms behind the emergence of 
stick-slip in boundary-lubricated systems have been numerically demonstrated to include 
repeated crystallization and shear melting of the thin lubricant film \cite{Thompson1990},
interlayer slips within the ordered solid-like lubricant film, or wall slips at the 
wall-film interface \cite{lei2011stick}. Most of the numerical studies of stick-slip
in boundary lubrication have focused on coarse-grained or simplified/idealized models
\cite{Thompson1990,oma,PhysRevLett.96.206102}, not explicitly considering the
atomic scale interactions occurring in real systems. On a coarse-grained scale, a useful 
classification of the lubricant-surface interactions is given by the wettability 
of the confining surfaces by the lubricant, with systems displaying a larger contact 
angle/lower wetting generally exhibiting lower friction. Other approaches to friction 
control include e.g. applying mechanical oscillations \cite{capozza2009suppression,giacco2012solid}.
While the effect of wettability on lubricated friction has been 
studied experimentally in macroscopic \cite{pawlak2011relationship,bhushan2008wetting,
kalin2013effect,borruto1998influence} and nanoscale \cite{mougin2004nanoscale} systems, 
and modeled using phenomenological finite-element models \cite{joly2009wetting} and 
simplified molecular dynamics (MD) simulations of nanopatterned surfaces 
\cite{cottin2003low,capozza2008lubricated}, less is known about the underlying atomic 
scale processes and mechanisms responsible for the presence or absence of stick-slip.

\begin{figure}[b!]
  \centering
  \includegraphics[trim=0cm 4cm 0cm 7cm, clip=true,width=0.38\textwidth]{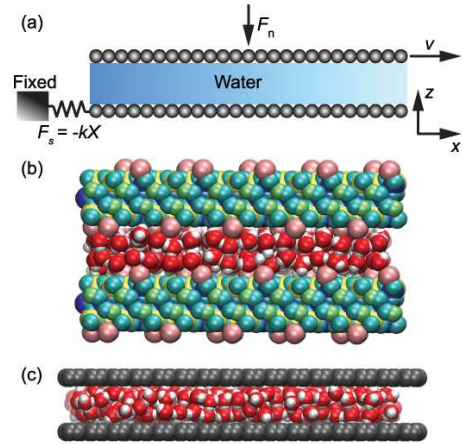}
  \caption{(color online) (\textbf{a}) The geometry of the simulation system. Solid sheets 
    are held together by a constant normal load $F_\text{n}$. The top sheet is 
    moving at a constant velocity $V$, and the bottom sheet is connected to a fixed 
    stage by a spring of stiffness $k$. The water molecules are confined by (\textbf{b}) two mica 
    sheets (each of thickness of 8.34 {\AA}) or (\textbf{c}) two monolayer graphene sheets. The 
    color code of the atoms is: water oxygen (red), water hydrogen (white), potassium 
    (pink), silicon (yellow), aluminum (blue), mica oxygen (cyan), mica hydrogen (lime), 
    and carbon (gray).}
  \label{system}
\end{figure}

Given the large surface-to-volume ratio in boundary lubrication, nature of the interaction 
between the lubricant and the confining surfaces originating from their atomic composition 
should play a crucial role. Thus, 
we study the interaction of a thin water layer (thickness $h$ around 
0.5 nm unless stated otherwise) in MD simulations using full atomic models of two 
experimentally relevant confining surfaces with different wetting characteristics: 
crystalline mica, a hydrophilic substrate that strongly adsorbs water \cite{Leng2006} and 
graphene, a hydrophobic surface interacting weakly with water \cite{leenaerts2009water}, 
see Fig. \ref{system}. We observe stick-slip dynamics for thin water layers confined by 
mica: each unit cell of mica contains two K$^+$ ions, interacting 
strongly with the water oxygens via Coulomb interactions, leading to formation of atomic 
scale capillary bridges next to the K$^+$ ions, connecting the two mica surfaces in the 
stick state. These bridges break during the subsequent slip event, and reform during the 
next stick phase, a process that is also visible as the breaking and reforming of 
interfacial hydrogen bonds between water and mica. This mechanism is different from both 
the crystallization-shear melting transitions \cite{Thompson1990} and interlayer or 
lubricant-surface slips \cite{lei2011stick} observed before in simplified models. In contrast, 
water films confined by hydrophobic graphene, as well as thicker water layers confined 
by mica, exhibit fundamentally different dynamics with no stick-slip.

To model the confined water film, we consider systems ranging from $200$ to $1200$ 
SPC/Fw water molecules \cite{Wu2006}. We consider 2M1-muscovite mica with the 
formula KAl$_2$(Al, Si$_3$)O$_{10}$(OH)$_2$, with 
the force field parameters from Ref. \cite{Heinz2005}. One mica surface consists of 
$10 \times 6 $ unit cells, and  has linear dimensions of $L_x = 52.07$ {\AA} and 
$L_y = 54.036$ {\AA}, see Fig. \ref{system}. To create site 
disorder, mimicking a real mica surface with a random distribution of potassium ions on 
it, one K$^+$ ion of the pair in each unit cell is removed and subsequently placed 
on the bottom part of the sheet \cite{Malani2009}. The graphene 
sheets have $L_x = 68.063$ {\AA} and $L_y = 36.841$ {\AA}. The 
Lennard-Jones parameters for carbon are from Ref. \cite{Saito2001}. 
The cutoff radius is $r_c = 10.0$ {\AA} for all potentials. Both sheets are parallel to 
the $xy$ plane with periodic boundary conditions along the $x$ and $y$ directions. 
Couette flow is generated by moving the top sheet at a constant velocity $V$ 
along the $x$ direction. The distance between parallel sheets is allowed to vary, and a 
constant normal load $F_\textrm{n}$, giving rise to a pressure $P_\perp$, is applied on the top 
sheet. The bottom sheet is constrained to move along the $x$ axis, and is attached to a spring 
of stiffness $k / N_{\textrm p}$ $=0.0035$ N/m, where $N_{\textrm p}$ is the total number of 
atoms in a sheet. The other end of the spring is connected to a fixed stage. 
Temperature of $T=295$ K is maintained using a Langevin 
thermostat, applied only in the $y$ direction to avoid streaming bias 
\cite{PhysRevA.41.6830, Thompson1997}. The equations of motion are solved with the velocity 
Verlet algorithm implemented in the LAMMPS code \cite{Plimpton1995}, with an integration 
time step of $1 \;$fs. Long-range electrostatic interactions are computed using the 
particle-particle particle-mesh (PPPM) solver with $10^{-5}$ accuracy. Initially the water 
molecules are arranged in a simple cubic lattice. The simulations are first run for 
$100 \; $ps with both surfaces kept fixed, followed by $100 \; $ps during which the top 
surface is subject to a normal force $F_\text{n}$ and is allowed to move vertically. 
Then, the top surface is driven horizontally with a velocity $V$ for $1 \; $ns to generate 
the steady state, after which  we continue the simulations for approximately $60 \; $ns, 
recording the observables of interest.

\begin{figure}[t!]
  \centering
  \includegraphics[trim=1.3cm 0.2cm 0cm 1cm, clip=true, width=0.4\textwidth]{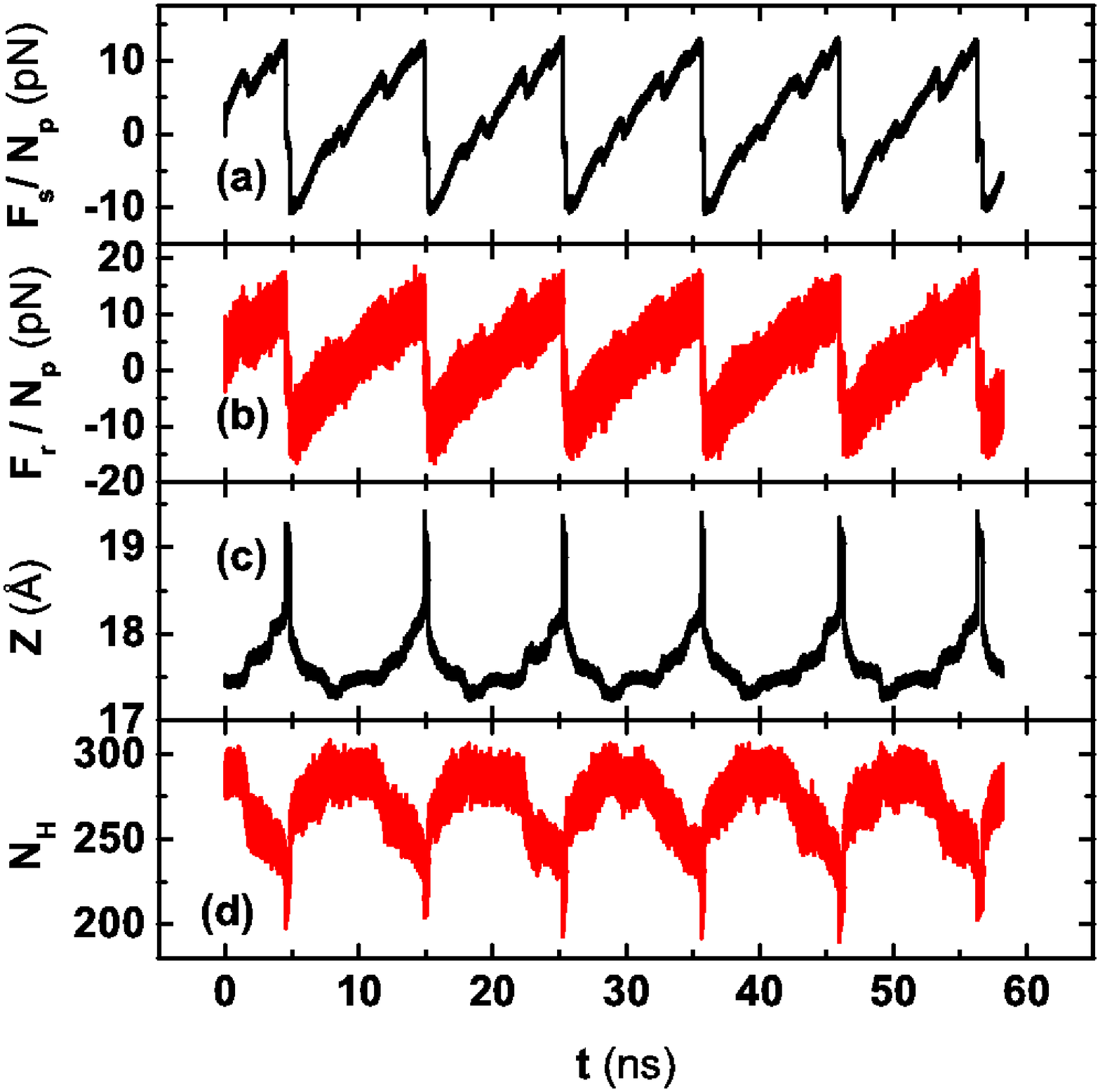}\\ 
  \includegraphics[trim=0.6cm 0.1cm 1.7cm 1.1cm, clip=true, width=0.43\textwidth]{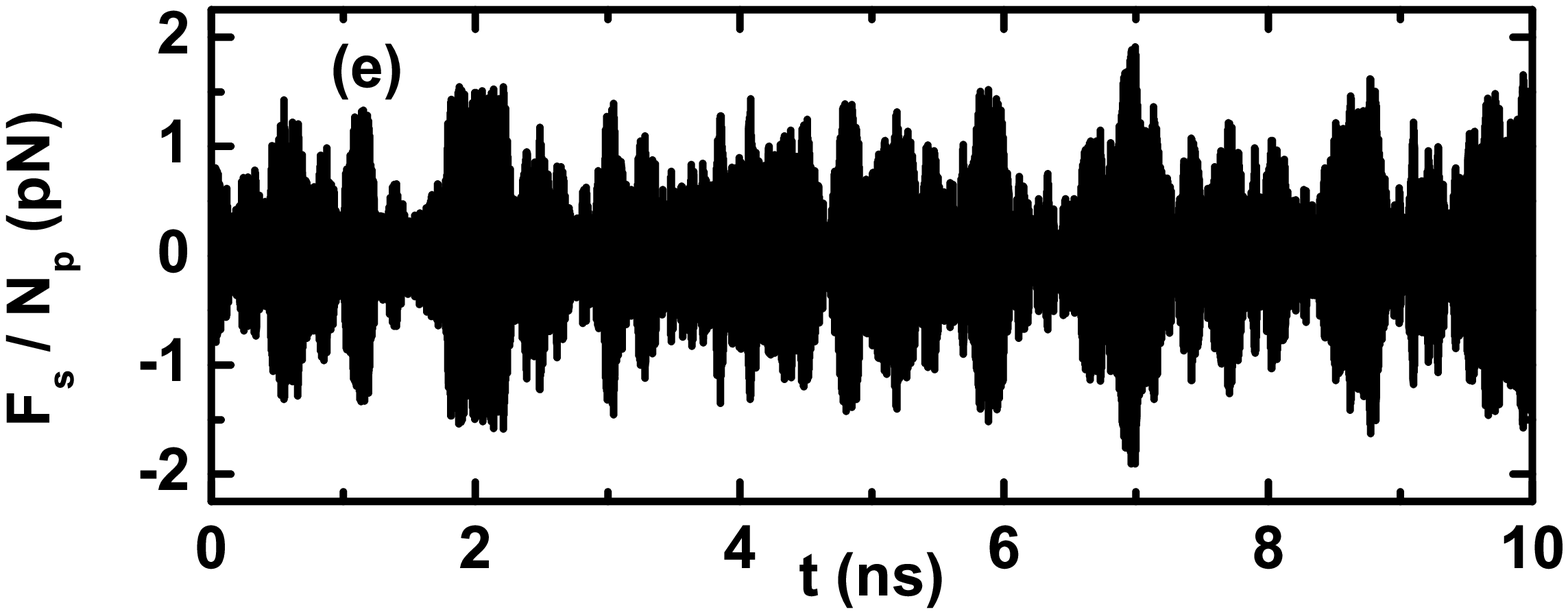} 
  \caption{(color online) Time evolution of (\textbf{a}) the spring force per sheet atom, (\textbf{b}) the 
    friction force per sheet atom on the bottom mica sheet applied by the water and the top mica 
    sheet, (\textbf{c}) the position $Z$ of the center of mass of the top mica sheet in the $z$ direction, 
    and (\textbf{d}) the number of hydrogen bonds between the 256 mica-confined water molecules 
    and the bottom mica sheet. (\textbf{e}) Spring force per sheet atom for 200 water molecules 
    confined by graphene sheets. $V = 0.1 \;$m/s and $P_\perp = 1 \;$atm in both cases.}
  \label{Mica_Stick_slip}
\end{figure}

Simulation results for $256$ water molecules confined by mica sheets for 
$P_\perp = 1 \;$atm and $V = 0.1 \;$ m/s are shown in Fig. \ref{Mica_Stick_slip}. The force 
per atom on the bottom sheet applied by the spring, $F_{\textrm s}/N_{\textrm p}$,
exhibits characteristic stick-slip behavior [Fig. \ref{Mica_Stick_slip} (a)]. 
Fig. \ref{Mica_Stick_slip} (b) shows the friction force per sheet atom on the bottom mica 
plate applied by the water and the top mica plate, $F_{\textrm r}$/$N_\textrm{p}$, 
exhibiting similar time-dependence as the spring force, 
with superimposed high-frequency fluctuations due to the finite temperature. 
Fig. \ref{Mica_Stick_slip} (c) shows the position $Z$ of the center of mass of the top 
sheet in the $z$ direction. The center of the bottom mica sheet is fixed at $z=4.16$ 
{\AA}. During each slip event, $Z$ increases by roughly $10 \%$ \cite{Thompson1990}. 
Since formation and breaking of interfacial chemical bonds is known to play a role in 
friction (see Ref. \cite{Li2011} for an example from rock friction), we show also 
the time-dependence of the number of hydrogen bonds (i.e. the number of water hydrogens 
closer than 3 {\AA} from the bottom mica surface) between water and the bottom mica 
surface in Fig. \ref{Mica_Stick_slip} (d): bonds break as the system evolves from 
stick to the slip state.  

For comparison, we also performed MD simulations of water confined by 
hydrophobic graphene sheets. 
We varied the number of water molecules from $200$ to $1200$, the normal loads from  
$P_\perp = 1$ to 10 atm,  and the driving velocities from 
$V = 0.01$ to 0.1 m/s. Fig. \ref{Mica_Stick_slip} (e) shows the spring 
force from simulations of $200$ water molecules, $P_\perp = 1 \;$atm, 
and $V = 0.1 \;$m/s; similar results are obtained for other $P_\perp$ and $V$ values. 
We observe a small increase of friction with $V$ for both mica and graphene, see 
Supplemental Material \cite{SM}, and Refs. \cite{chen2006velocity,ohnishi2002humidity}
for experimental results on mica-confined systems with sliding velocities
significantly lower than those reachable in our MD simulations. 
Our simulations thus demonstrate that the stick-slip behavior 
does not arise for thin water films confined by graphene. Instead, continuous, smooth
sliding with the maximum friction force well below that obtained for mica is observed for
all parameter values considered. We also note that the same applies to the mixed
system with one graphene and one mica surface: slip is localized at the hydrophobic 
graphene-water interface, and no stick-slip is observed.

This difference between the two kinds of surfaces may be explained by the relatively strong 
interaction of the potassium ions on the mica surfaces with the oxygen atoms of the 
water molecules via Coulomb interactions. Thus, the ions could act as \lq\lq freezing 
nuclei\rq\rq, with the water molecules gathering around them to form nanoscale capillary 
water bridges \cite{PhysRevLett.88.185505, PhysRevLett.95.135502}, connecting the 
top and bottom surfaces within the stick phase. As the system starts to slip, these bridges 
would break. The interaction of carbon atoms with oxygen is much 
weaker, and we expect that no capillary bridges are formed between graphene sheets, 
explaining the absence of stick-slip dynamics in that case. 

\begin{figure}[t!]
  \centering
  \includegraphics[trim=2cm 0.3cm 0cm 1.7cm, clip=true, width=0.5\textwidth]{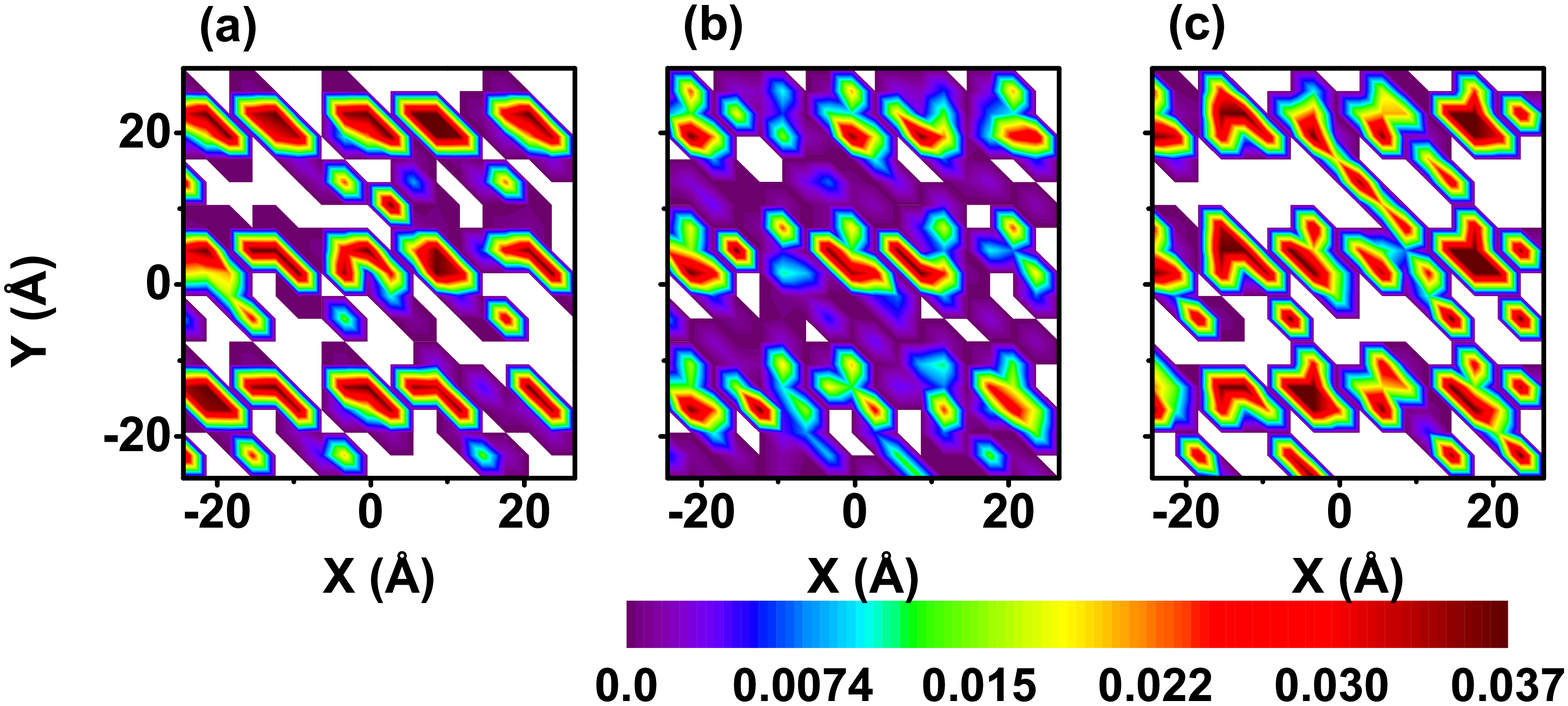}\\ 
  \includegraphics[trim=0cm 0.25cm 0cm 1.25cm, clip=true, width=0.4\textwidth]{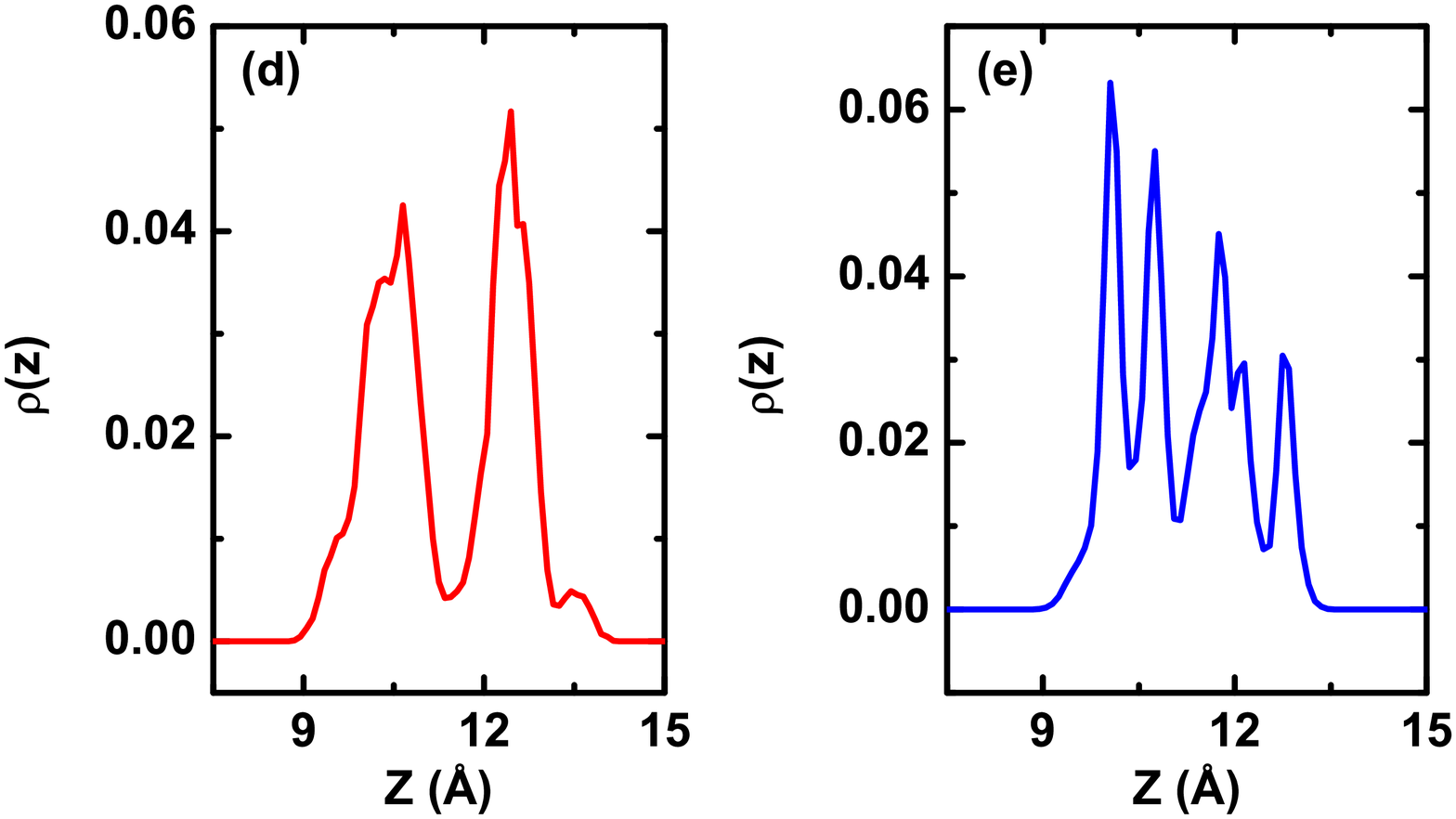} 
  \caption{(color online) Contour graphs of the density distribution $\rho(x,y)$ 
    of water oxygens in the contact layer relative to the bottom mica surface for 
    (\textbf{a}) $t = 1 \;$ns (\lq\lq stick\rq\rq), (\textbf{b}) $t = 5 \;$ns 
    (\lq\lq slip\rq\rq), and (\textbf{c}) $t = 7 \;$ns (\lq\lq stick\rq\rq). 
    White corresponds to no water molecules being present. The density profiles 
    across the gap, $\rho(z)$, of water confined by mica sheets when the 
    system (\textbf{d}) slips and (\textbf{e}) is in the stick state. In both (\textbf{d})
    and (\textbf{e}), the top surface of the bottom mica sheet is at $z = 8.3$ {\AA},
    while the lower surface of the top mica sheet is at $z = 14.7$ {\AA} in (\textbf{d})
    and at $z = 13.4$ {\AA} in (\textbf{e}).}
  \label{Mica_Oxygen_Distribution}
\end{figure}

To verify this hypothesis, we calculate the density distributions $\rho(x,y)$ of 
water oxygens in the contact layer relative to the bottom surfaces. Fig. 
\ref{Mica_Oxygen_Distribution} (a) shows $\rho(x,y)$ for a water film confined 
by mica sheets when the system sticks [$t = 1 \;$ns in Fig. \ref{Mica_Stick_slip} (a)]. 
Peaks in $\rho(x,y)$ are located at the K$^+$ ions. Fig. \ref{Mica_Oxygen_Distribution} (b) 
presents the corresponding $\rho(x,y)$ during the first slip state when $t = 5 \;$ns [cf. again
Fig. \ref{Mica_Stick_slip} (a)]: the peaks of $\rho(x,y)$ become smaller and broader. Finally, Fig. 
\ref{Mica_Oxygen_Distribution} (c) shows $\rho(x,y)$ for the subsequent stick state at 
$t = 7 \;$ns [Fig. \ref{Mica_Stick_slip} (a)], where we again observe that the peaks are 
as high and narrow as those of the previous stick state.

\begin{figure}[t!]
  \centering
  \includegraphics[trim=0.25cm 1.75cm 0cm 1.5cm, clip=true, width=0.5\textwidth]{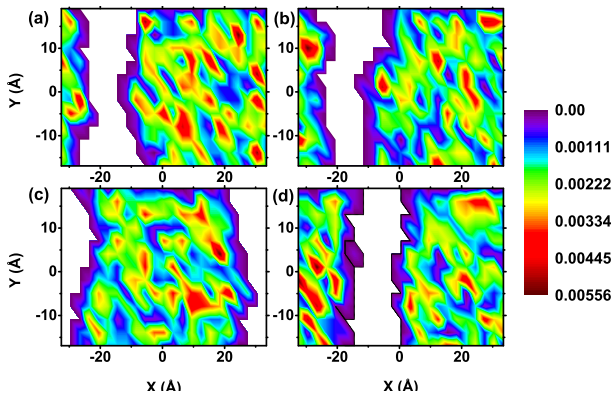} 
  \caption{(color online) Contour graphs of the density distribution $\rho(x,y)$ 
    of water oxygens in the contact layer relative to the bottom graphene surface for 
    (\textbf{a}) $t = 0 \;$ns, (\textbf{b}) $t = 3 \;$ns, (\textbf{c}) $t = 5 \; $ns 
    and (\textbf{c}) $t = 8 \;$ns. White corresponds to no water molecules being present.}
  \label{Graphene_Oxygen_Distribution}
\end{figure}

To gain more insight into the nucleation and breaking of the capillary bridges between the 
surfaces, we calculate the density profiles $\rho(z)$ of water oxygens across the gap. 
When the system is slipping [Fig. 
\ref{Mica_Oxygen_Distribution} (d)], $\rho(z)$ exhibits two separate peaks, 
consistent with breaking of the capillary bridges. In the stick state [Fig. 
\ref{Mica_Oxygen_Distribution} (e)], $\rho(z)$ exhibits multiple peaks spanning 
the gap. This can be understood as the water molecules forming nanoscale capillary bridges 
between the two mica surfaces. In contrast to this behavior, the density distributions $\rho(x,y)$ 
of the water film consisting of $200$ water molecules confined by graphene sheets in 
Fig. \ref{Graphene_Oxygen_Distribution} show that water clusters to form a single, relatively
large droplet-like structure between the two graphene sheets, without any apparent 
signature of breaking-reforming transitions. The corresponding density profiles
$\rho(z)$ \cite{SM} are similar to previous observations in equilibrium
graphene-confined systems \cite{cicero2008water}.

\begin{figure}[t!]
\centering
\includegraphics[trim=1.25cm 1.35cm 0cm 1.75cm, clip=true, width=0.5\textwidth]{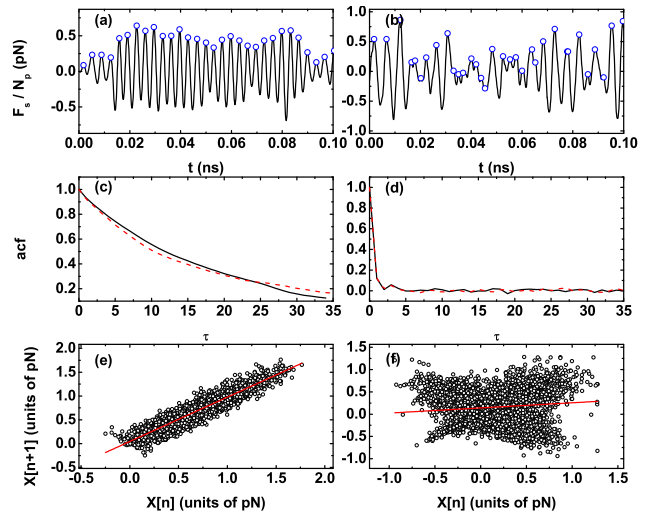} 
\caption{
  (color online) Time dependence of $F_\text{s}/N_\text{p}$ during a time period of 0.1 ns 
  ({\bf a}) for 1200 graphene-confined and ({\bf b}) for 1536 mica-confined water molecules. 
  Blue circles show the local maxima of the signals, corresponding
  to the time-varying amplitudes $X(n)$ of the spring force oscillations.
  The autocorrelation functions (as a function of the lag $\tau \equiv n-n'$) of these 
  amplitudes are given ({\bf c}) for graphene and ({\bf d}) for mica, extracted 
  from $10 \; $ns long spring force signals. The solid lines correspond to the simulation 
  results, while the dashed lines show the corresponding ACFs from the autoregressive model. 
  The plots of $X[n+1]$ vs $X[n]$ extracted from the simulations for ({\bf e}) graphene and 
  ({\bf f}) mica further illustrate the different nature of the smooth sliding dynamics for 
  the two kinds of confining surfaces. The slopes of the lines (linear fits) are 0.9 and 0.1 
  for graphene and mica, respectively.}
\label{ACF}
\end{figure}

Thus, when the two mica surfaces are very close together, the thin confined water film 
loses its fluidity, and the bulk flow properties of water play little or no role in 
friction. However, they may be recovered by increasing the thickness of the water layer 
\cite{PhysRevLett.94.026101}, with the conditions approaching those of hydrodynamic 
lubrication. To this end, we performed MD simulations with four different, larger 
thicknesses of the water layer: $h=1.77$, 2.03, 2.29, and 2.56 nm, 
corresponding to 1536, 1792, 2048, and 2304 water molecules, respectively. For these 
thicker water films, the stick-slip dynamics disappears. Instead, smooth 
sliding dynamics is observed, which at a first glance looks similar to that in the 
graphene-confined system. However, subtle differences can still be observed between the 
two surfaces. Zooming in to the spring force time series (e.g. the one shown 
in Fig. \ref{Mica_Stick_slip}(e)] reveals periodic oscillations corresponding to the eigenfrequencies 
of the spring-bottom plate mass ($M$) system, $f=1/(2\pi) \sqrt{k/M}$. For both 
surfaces, the varying amplitudes of these oscillations at each period [blue 
circles in Figs. \ref{ACF}(a) and (b)] form sequences of time-ordered observations $X(n)$ 
which can be well-described by an autoregressive model $X(n+1) = \alpha X(n) + W(n)$
[or equivalently, the Ornstein-Uhlenbeck process $X(n+1)-X(n) = -(1-\alpha)X(n) + W$], with 
$W$ white noise originating from the interaction with the fluctuating lubricant and 
$\alpha$ a model parameter, both extracted using the R package \cite{RDevelopmentCoreTeam2011}. 
For water confined by graphene, we find $\alpha \approx 0.8$ and $\delta W \approx 0.1$ pN
for all conditions considered, while we find $\alpha \approx 0.1$ and $\delta W \approx 0.3$  
pN for thick water films ($h \geq 1.77$ nm) confined by mica. Accordingly, the 
autocorrelation function (ACF) of $X(n)$ for mica decays more rapidly to zero than its 
counterpart for graphene. In both cases the ACFs computed from the simulation data agree 
with those of the corresponding autoregressive model [see Figs. \ref{ACF} (c) and (d)]. 
The observation that $\delta W$ does not significantly depend on $h$ for $h \geq 1.77$ nm 
indicates that the screened mica-water interaction has a sub-nanometer range, resulting 
essentially in a surface effect of the fluctuations of the water layer. Also, the stronger 
interaction of mica with the fluctuating lubricant (as compared to that of 
graphene) results in a factor of three larger $\delta W$ [see also Figs. 
\ref{ACF} (e) and (f)].

In summary, the presence or absence of breaking-reforming transitions of local capillary bridges 
in the water film, controlled by the atomic structure and the ensuing wettability (hydrophilic mica
vs hydrophobic graphene) of the confining surfaces, plays a crucial role in whether stick-slip 
dynamics is observed or not. For mica, the decisive role of the K$^+$ ions in the formation of the 
nanoscale capillary bridges suggest that the microscopic details behind stick-slip
dynamics should in general depend on the atomic structure of the system, and it would be interesting 
to perform similar studies for other confining surfaces with different surface-lubricant interactions. 
Nevertheless, we expect our main observations to be rather general, and to open up interesting 
possibilities in controlling nanoscale boundary-lubricated friction by tuning the wettability 
of the confining surfaces.

\begin{acknowledgments}
We acknowledge the financial support by the Academy of Finland through the Centres 
of Excellence Program (project no. 251748) and via an Academy Research Fellowship
(L.L., project no. 268302). We are also grateful to the COST action MP1303. The 
calculations presented above were performed using computer resources within the Aalto 
University School of Science \lq \lq Science-IT\rq \rq project. We also acknowledge the 
computational resources provided by CSC (Finland).
\end{acknowledgments}


%

\end{document}